\documentclass[twocolumn,showpacs,nofootinbib,10pt]{revtex4}

\usepackage{color}
\usepackage{graphicx}
\usepackage{graphicx,float}\usepackage[all]{xy}
\usepackage{amsmath,upgreek,tikz,bm}
 \newcommand{\blt}{\textcolor{black}}
\usepackage{amssymb}

\usepackage{textgreek}

\newcommand{\beq}{\begin{eqnarray}}
\newcommand{\eeq}{\end{eqnarray}}
\newcommand{\be}{\begin{equation}}
\newcommand{\ee}{\end{equation}}

\newcommand{\bea}{\begin{eqnarray}}
\newcommand{\eea}{\end{eqnarray}}
\newcommand{\bes}{\begin{subequations}}
\newcommand{\ees}{\end{subequations}}

\newcommand{\ba}{\begin{eqnarray}}
\newcommand{\ea}{\end{eqnarray}}
\newcommand\orcidroldao{{\href{https://orcid.org/0000-0003-3978-532X}{\orcidicon}}}
\newcommand{\orcidicon}{%
	\begin{tikzpicture}
	\draw[lime, fill=lime] (0,0)
		circle [radius=0.16]
		node[white] {{\fontfamily{qag}\selectfont \tiny ID}};
	\draw[white, fill=white] (-0.0625,0.095)
		circle [radius=0.007];
	\end{tikzpicture}	\hspace{-2mm}
}

\usepackage[colorlinks,hyperindex,unicode]{hyperref}
\definecolor{green1}{RGB}{0,128,0} 
\hypersetup{hidelinks,backref=true,pagebackref=true,hyperindex=true,colorlinks=true,breaklinks=true,urlcolor= blue}
\hypersetup{%
  colorlinks = true,
  linkcolor  = blue,
  citecolor = cyan,
}
\usepackage{bookmark,textgreek}
\usepackage{hyperref,color,xcolor}
\hypersetup{hidelinks,hyperindex=true,colorlinks=true,breaklinks=true,urlcolor= blue}
\hypersetup{%
  colorlinks = true,
  linkcolor  = blue
}
\begin{document}
\title{Gravitational decoupling, hairy black holes and conformal anomalies}

\author{Pedro Meert}
\email{pedro.meert@ufabc.edu.br} 
\affiliation{Center of Physics, Universidade Federal do ABC,  09210-580, Santo Andr\'e, Brazil.}
\author{Roldao da Rocha\orcidroldao\!\!}
\email{roldao.rocha@ufabc.edu.br}
\affiliation{Center of Mathematics,  Federal University of ABC, 09210-580, Santo Andr\'e, Brazil.}

%
%


\begin{abstract} 
Hairy black holes in the gravitational decoupling setup are studied from the perspective of conformal anomalies. Fluctuations of decoupled sources can be computed by measuring the way the trace anomaly-to-holographic Weyl anomaly ratio differs from the unit. Therefore the gravitational decoupling parameter governing three hairy black hole metrics is then bounded to a range wherein one can reliably emulate AdS/CFT with gravitational decoupled solutions, in the tensor vacuum regime.

\end{abstract}



\maketitle

\section{Introduction} 

Gravitational decoupling methods comprise established successful protocols used to generate analytical solutions of the Einstein's effective field equations \cite{Ovalle:2017fgl,Ovalle:2019qyi,Ovalle:2020kpd,Casadio:2012rf,Ovalle:2017wqi,Antoniadis:1998ig}. 
 The gravitational decoupling and some extensions were studied in Refs. \cite{covalle2,Ovalle:2014uwa,Ovalle:2016pwp,Casadio:2013uma,Ovalle:2013vna,Ovalle:2013xla,Ovalle:2018vmg,Casadio:2012pu,Casadio:2015jva,Casadio:2016aum} and have been applied to kernel solutions of general relativity to construct new physically realistic solutions that describe stellar distributions, including anisotropic ones \cite{daRocha:2020rda,Fernandes-Silva:2017nec,Contreras:2018gzd,Ovalle:2007bn,Sharif:2018tiz,Sharif:2019mzv,Morales:2018urp,Rincon:2019jal,Hensh:2019rtb,Ovalle:2019lbs,Gabbanelli:2019txr,Tello-Ortiz2020,Gabbanelli:2018bhs,Panotopoulos:2018law,Heras:2018cpz,Contreras:2018vph,Tello-Ortiz:2020euy}. Refs. \cite{Fernandes-Silva:2019fez,daRocha:2017cxu,Fernandes-Silva:2018abr} derived accurate physical constraints on the parameters in gravitational decoupled solutions, using the WMAP, eLISA and LIGO.  
The gravitational decoupling procedure iteratively constructs, upon a given isotropic source of gravitational field, anisotropic compact sources of gravity, that are weakly coupled. One starts with a perfect fluid, then coupling it to more elaborated stress-energy-momentum tensors that underlie realistic compact configurations \cite{Maurya:2019kzu,PerezGraterol:2018eut,Morales:2018nmq,Contreras:2019iwm,Contreras:2019fbk,Singh:2019ktp,Tello-Ortiz:2019gcl,Maurya:2019xcx,Cedeno:2019qkf,Sharif:2018toc,Estrada:2018zbh,Torres:2019mee,Abellan:2020jjl,Estrada:2018vrl,Leon:2019abq,Casadio:2019usg,Sharif:2019mjn,Abellan:2020wjw,Rincon:2020izv,Sharif:2020arn,Maurya:2020gjw}.

Any action related to a classical conformal theory is invariant under Weyl transformations. Since the variation of the action with respect to the background metric is proportional to the stress-energy-momentum tensor, then the variation of the action with respect to a conformal rescaling is proportional to the trace of the stress-energy-momentum tensor, which vanishes for conformally invariant theories. However, upon quantization, conformal invariance under Weyl rescalings may be broken and conformal anomalies set in \cite{Capper:1973mv}. In this case, the trace of the stress-energy-momentum tensor may achieve a non-null expectation value and, thus, a conformal anomaly regards a trace anomaly \cite{Duff:1993wm,Henningson:1998gx,Kuntz:2017pjd,Bonora:1983ff,Bonora:2014qla,Kuntz:2019omq}.
In the context of the gravitational decoupling procedure, comparing the holographic Weyl anomaly to the trace anomaly of the energy-momentum tensor from 4D field theory leads to a quantity that can probe and measure the source of the gravitational decoupling \cite{Casadio:2003jc}. Hence, the calculation of the trace anomaly-to-holographic Weyl anomaly ratio makes one capable to place the gravitational decoupling, in the context of three possible metrics describing hairy black holes, as a reliable AdS/CFT realization.
 
This paper is organized as follows: Sec. \ref{Sgd} is dedicated to reviewing the gravitational decoupling procedure, obtaining three different metrics for gravitational decoupled hairy black holes. In Sec. \ref{dsfb}, the trace anomalies are computed for these three solutions, from the point of view of CFT, and compared to the respective values predicted by the AdS/CFT duality. Sec. \ref{4} is dedicated to conclusions.

\section{Gravitational decoupling and hairy black holes}
\label{Sgd}
The gravitational decoupling procedure can be straightforwardly introduced when kernel solutions of Einstein's effective field equations can be used to decouple any intricate stress-energy-momentum
tensor into manageable pieces \cite{Ovalle:2017wqi,Ovalle:2019qyi}, including the case of hairy black holes \cite{Ovalle:2020kpd}. When one regards Einstein's field equations,
\begin{equation}
\label{corr2}
G_{\mu\nu}:=
R_{\mu\nu}-\frac{1}{2}R g_{\mu\nu}
=
\upkappa^2\,\mathring{T}_{\mu\nu},
\end{equation}
where the stress-energy-momentum tensor, satisfying the conservation equation $
\nabla_\mu\,\mathring{T}^{\mu\nu}=0$, can be split as 
\begin{equation}
\label{emt}
\mathring{T}_{\mu\nu}
=
\mathsf{T}^{\rm}_{\mu\nu}
+
\upalpha\,\Uptheta_{\mu\nu},
\end{equation}
for $\mathsf{T}_{\mu\nu}$ being a general-relativistic solution and $\Uptheta_{\mu\nu}$ encoding additional sources in the gravitational sector, for $\upalpha$ being an arbitrary decoupling parameter that is not perturbative, in general.  
One considers static, spherically symmetric, stellar distributions described by the metric  
\begin{equation}
ds^{2}
=
e^{\upnu (r)}dt^{2}-e^{\uplambda (r)}dr^{2}
-r^{2}d\Omega^2,
\label{metric}
\end{equation}
where $d\Omega^2$ denotes the solid angle element.
The Einstein's field equations~(\ref{corr2}) are equivalently written as \bes
\begin{eqnarray}
\label{ec1}
\!\!\!\!\!\!\!\!\!\!\!\!\upkappa^2\!
\left(
\mathsf{T}_0^{\ 0}+\Uptheta_0^{\ 0}
\right)
&\!=\!&
\frac 1{r^2}
-
e^{-\uplambda }\left( \frac1{r^2}-\frac{\uplambda'}r\right),
\\
\label{ec2}
\!\!\!\!\!\!\!\!\upkappa^2\!
\left(\mathsf{T}_1^{\ 1}+\Uptheta_1^{\ 1}\right)
&\!=\!&
\frac 1{r^2}
-
e^{-\uplambda }\left( \frac 1{r^2}+\frac{\upnu'}r\right),
\\
\label{ec3}
\!\!\!\!\!\!\!\!\!\!\!\!\!\!\!\upkappa^2\!
\left(\mathsf{T}_2^{\ 2}\!+\!\Uptheta_2^{\ 2}\right)
&\!=\!&
\!-\frac {e^{-\uplambda }}{4}
\left(2\upnu''\!+\!\upnu'^2\!-\!\uplambda'\upnu'
\!+\!2\,\frac{\upnu'\!-\!\uplambda'}r\right)
\end{eqnarray}
\ees
where the prime denotes the derivative with respect to the variable $r$.
Eqs. (\ref{ec1} -- \ref{ec3}) regard the effective density, and the effective radial and tangential pressures, respectively given by \cite{Ovalle:2017wqi,Ovalle:2019qyi}
\blt{\bes
\beq
\mathring{\rho}
&=&
\rho+
\upalpha\Uptheta_0^{\ 0},\label{efecden}\\
\mathring{p}_{r}
&=&
p
-\upalpha\Uptheta_1^{\ 1},
\label{efecprera}\\
\mathring{p}_{t}
&=&
p
-\upalpha\Uptheta_2^{\ 2}, 
\label{efecpretan}
\eeq\ees} \!with anisotropy \beq
\Updelta = 
\mathring{p}_{t}-\mathring{p}_{r}.\eeq

A solution to Einstein's field equations \eqref{corr2} for the single kernel source $\mathsf{T}_{\mu\nu}$ was considered \cite{Ovalle:2017wqi,Ovalle:2020kpd}, 
\begin{equation}
ds^{2}
=
e^{\upxi (r)}dt^{2}
-e^{\upmu (r)}dr^{2}
-
r^{2}d\Omega^2
,
\label{pfmetric}
\end{equation}
where 
\begin{equation}
\label{standardGR}
e^{-\upmu(r)}
\equiv
1-\frac{\upkappa^2}{r}\int_0^r x^2\,\mathsf{T}_0^{\ 0}(x)\, dx
=
1-\frac{2m(r)}{r}
\end{equation}
is the Misner--Sharp--Hernandez function.
The additional source $\Uptheta_{\mu\nu}$ drives the gravitational decoupling of the kernel metric~\eqref{pfmetric}, implemented by the mappings 
\bes
\begin{eqnarray}
\label{gd1}
\upxi(r)
&\mapsto &
\upnu(r)=\upxi(r)+\upalpha g(r)
\\
\label{gd2}
e^{-\upmu(r)} 
&\mapsto &
e^{-\uplambda(r)}=e^{-\upmu(r)}+\upalpha f(r)
, 
\end{eqnarray}
\ees
where $f(r)$ [$g(r)$] is the geometric deformation for the radial [temporal] metric
component.
Eqs.~(\ref{gd1}, \ref{gd2}) split the Einstein's field equations~(\ref{ec1}) -- (\ref{ec3}) into two distinct arrays. 
The first one encodes the Einstein's field equations for $\mathsf{T}_{\mu\nu}$, solved by the kernel metric~(\ref{pfmetric}). The second one is associated to $\Uptheta_{\mu\nu}$ and reads
\bes
\begin{eqnarray}
\label{ec1d}
\!\!\!\!\!\upkappa^2\,\Uptheta_0^{\ 0}
&=&
-\upalpha\left(\frac{f}{r^2}+\frac{f'}{r}\right),
\\
\label{ec2d}
\!\!\!\!\!\upkappa^2\,\Uptheta_1^{\ 1}
+\upalpha\,\frac{e^{-\upmu}\,g'}{r}
&\!=\!&
-\upalpha\,f\left(\frac{1}{r^2}+\frac{\upnu'}{r}\right)
\\
\label{ec3d}
\!\!\!\!\!\!\!\!\!\!\!\!\!\upkappa^2\Uptheta_2^{\ 2}\!+\!\upalpha{f}\left(2\upnu''\!+\!\upnu'^2\!+\!\frac{2\upnu'}{r}\right)\!&\!=\!&\!-\upalpha\frac{f'}{4}\!\left(\upnu'\!+\!\frac{2}{r}\right)\!+\!V
\end{eqnarray}
\ees
where \cite{Ovalle:2017fgl}
\beq
V(r) = \upalpha e^{-\upmu}\left(2g''+g'^2+\frac{2\,g'}{r}+2\upxi'\,g'-\upmu'g'\right)\eeq 
The tensor-vacuum, defined for $\Uptheta_{\mu\nu}\neq 0$ and $\mathsf{T}_{\mu\nu}=0$, leads to hairy black hole solutions \cite{Ovalle:2018umz}. 
Eqs. (\ref{ec1}) -- (\ref{ec2}) then yield a negative radial pressure, 
\begin{equation}
\mathring{p}_{r}
=
-\mathring{\rho}.
\label{schwcon}
\end{equation}
and, together to the Schwarzschild solution, it implies that 
\begin{equation}
\label{fg}
\upalpha\,f(r)
=
\left(1-\frac{2M}{r}\right)\left(e^{\upalpha\,g(r)}-1\right)
,
\end{equation}
so that the line element~\eqref{metric} becomes
\begin{eqnarray}
\label{hairyBH}
ds^{2}
&\!=\!&
\left(1-\frac{2M}{r}\right)
e^{\upalpha g(r)}
dt^{2}
\!-\!\left(1-\frac{2M}{r}\right)^{-1}
e^{-\upalpha\,g(r)}
dr^2
\nonumber
\\
&&
\qquad\qquad\qquad\qquad\qquad\qquad\qquad-r^{2}\,d\Omega^2
.
\end{eqnarray}
In the radial range $r\geq 2M$, the tensor-vacuum \blt{is given by expressing $\Uptheta_0^{\ 0}$ by the most general linear combination of the radial and tangential components of the stress-energy-momentum tensor, as 
\beq
\Uptheta_0^{\ 0}
=
a\,\Uptheta_1^{\ 1}+b\,\Uptheta_2^{\ 2},\eeq 
with $a,b\in\mathbb{R}$ denoting the coefficients of the linear combination}. 
Eqs.~(\ref{ec1d}) -- (\ref{ec3d}) then yield  
\begin{eqnarray}
\label{master}
\!\!\!\!\!\!\!\!\!\!\!\!\!\!\!\!\!\!&&b\,r\,(r-2M)\,h''+2\,\left[(a+b-1)\,r-2\,(a-1)\,M\right]
h'
\nonumber
\\
\!\!\!\!\!\!\!\!\!\!\!\!\!\!\!&&
+2\,(a-1)\,h=2\,(a-1)
,
\end{eqnarray}
for 
$ 
h(r)
=
e^{\upalpha\,g(r)}$. \blt{A trivial deformation corresponding to the standard Schwarzschild solution can be yielded when $a = 1$.}
The solution \blt{of Eq. (\ref{master})} can be written as
\begin{equation}
\label{master2}
e^{\upalpha\,g(r)}
=
1+\frac{1}{r-2M}
\left[\ell_0+r\left(\frac{\ell}{r}\right)^{n}
\right]
,
\end{equation}
where $\ell_0=\upalpha\ell$ is a primary hair charge, whereas \beq\label{nnn}
n
=
2\left(a-1\right)/b,\eeq
with $n>1$ for asymptotic flatness. 

\blt{In the tensor-vacuum background, this line element is produced by the effective density, the radial, and tangential pressures, respectively, }
\bes
\beq
\mathring{\rho}
&=&
\Uptheta_0^{\ 0}
=
\upalpha\,\frac{(n-1)\,\ell^n}{\upkappa^2\,r^{n+2}},
\label{efecdenx}\\
\mathring{p}_{r}
&=&
-\Uptheta_1^{\ 1}
=
-\mathring{\rho},\\
\mathring{p}_{t}
&=&
-\Uptheta_2^{\ 2}
=
\frac{n}{2}\,\mathring{\rho}
.
\label{efecptanx}
\eeq
\ees
On the other hand, the dominant energy conditions, \beq
\mathring{\rho}\geq |\mathring{p}_r|,\quad \mathring{\rho}\geq|\mathring{p}_t|,\label{dec0}\eeq yield $n\le2$ \cite{Ovalle:2017wqi,Ovalle:2019qyi,Ovalle:2020kpd}.
 Besides, the strong energy conditions, 
 \bes
\begin{eqnarray}
\mathring{\rho}+\mathring{p}_r+2\,\mathring{p}_t
\geq
0, 
\label{strong01}\\
\mathring{\rho}+\mathring{p}_r
\geq
0,\\
\mathring{\rho}+\mathring{p}_t
\geq
0,
\end{eqnarray}
\ees
\blt{make Eq.~\eqref{schwcon} to read 
\beq\label{t001}
-\Uptheta_0^{\ 0}\leq\Uptheta_2^{\ 2}\leq0.\eeq Therefore, together with Eqs.~\eqref{ec1d} and~\eqref{ec3d}, Eq. (\ref{t001}) can be written as} \bes\begin{eqnarray}
\label{strong5}
\!\!\!\!\!\!\!\!\!\!\!\!\!\!\!\!\!\!G_1(r)&:=&{h''(r-2M)+2h'}
\geq
0,\\
\label{strong51}
\!\!\!\!\!\!\!\!\!\!\!\!\!\!\!\!\!\!G_2(r)&:=&
h''r(r-2M)
+4h'M
-2h+2
\geq
0.
\end{eqnarray} \ees
The mapping 
\begin{equation}
\label{gauge}
h(r)
\mapsto
h(r)-\frac{\ell_0}{r-2M}
\end{equation}
leaves $G_1(r)$ and $G_2(r)$ invariant. 
Solutions with a proper horizon at $r\sim 2M$, which also
behave approximately like the Schwarzschild metric for $r\gg 2M$, yield $
G_1(r)=0$. Hence, solving Eq.~\eqref{strong5} implies that  
\begin{equation}
\label{strongg}
h(r)
=
c_1
-
\upalpha\,\frac{\ell-r\,e^{-r/M}}{r-2M}.
\end{equation}
Also, Eq.~\eqref{strongg} is also contrained to \eqref{strong51}. Replacing~\eqref{strongg} in ~\eqref{hairyBH} implies the metric 
\begin{equation}
\label{strongBH}
e^{\upnu}
=
e^{-\uplambda}
=
1-\frac{2{\cal M}}{r}+\upalpha\,e^{-r/({\cal M}-\upalpha\,\ell/2)}
,
\end{equation}
to represent a hairy black hole, 
where ${\cal M}=M+\upalpha\,\ell/2$.

Now, the strong energy conditions are consistent with $\ell
\geq
2M/e^{2}$, whose extremal case $\ell=2M/e^{2}$ leads to 
\begin{eqnarray}
\label{strongh2M}
e^{\upnu}
=
e^{-\uplambda}
=
1-\frac{2M}{r}+\upalpha\,\left(e^{-r/M}-\frac{2M}{e^2\,r}\right)
.
\end{eqnarray}
which has the horizon at $r_{\scalebox{.56}{\textsc{hor}}} = 2M$. 
The dominant energy conditions, 
\bes
\begin{eqnarray}
\mathring{\rho}
\geq
|\mathring{p}_r|,\\
\mathring{\rho}
\geq
|\mathring{p}_t|,
\label{dom2} 
\end{eqnarray}
\ees
 in terms of \eqref{efecden} and~\eqref{efecpretan} are respectively equivalent to 
\bes
\begin{eqnarray}
\label{dom6}
-r(r-2M)h''
-4(r-M)h'
-2h+2
&\geq&
0,
\qquad
\\
\label{dom61}
r\,(r-2M)\,h''
+4\,M\,h'
-2\,h+2
&\geq&
0.
\end{eqnarray} 
\ees
Solving~\eqref{dom6} for $r\sim
2M$ and $r
\gg
M$ yields \cite{Ovalle:2018umz} \begin{equation}
\label{dominantg}
h(r)
=
1
-
\frac{1}{r-2M}
\left(
\upalpha\,\ell
+\upalpha\,M\,e^{-r/M}
-\frac{Q^2}{r}
\right)
,
\end{equation}
where the charge $Q=Q(\upalpha)$ encompasses also tidal charges generated by additional gravitational sectors.
Eq.~\eqref{dominantg} also has to satisfy \eqref{dom61},
which reads
\begin{equation}
\frac{4\,Q^2}{r^2}
\ge
\frac{\upalpha}{M}\,(r+2M)\,e^{-r/M}
.
\end{equation}
Using~\eqref{dominantg} into the line element~\eqref{hairyBH}, yields \begin{equation}
\label{dominantBH}
e^{\upnu}
=
e^{-\uplambda}
=1-\frac{2M+\upalpha\,\ell}{r}
+\frac{Q^2}{r^2}
-\frac{\upalpha\,M\,e^{-r/M}}{r},
\end{equation}
such that 
\begin{equation}
\mathring{\rho}
=
\Uptheta_0^{\ 0}
=
-\mathring{p}_r
=
\frac{Q^2}{\upkappa^2\,r^4}
-\frac{\upalpha\,e^{-r/M}}{\upkappa^2\,r^2}
\label{dendom}
\end{equation}
The metric~\eqref{dominantBH} also represents hairy black holes, where $Q$ and $\ell_0=\upalpha\,\ell$
comprise charges generating primary hair.
\par
The horizon radii $r_{\scalebox{.56}{\textsc{hor}}}$ are given by solutions of
\begin{equation}
\label{horizondom}
\upalpha\,\ell
=
r_{\scalebox{.56}{\textsc{hor}}}
-2M
+\frac{Q^2}{r_{\scalebox{.56}{\textsc{hor}}}}
-\upalpha\,M\,e^{-r_{\scalebox{.56}{\textsc{hor}}}/M}
,
\end{equation}
which allows us to write the metric functions~\eqref{dominantBH} as
\begin{eqnarray}
\label{dominantBHH}
e^{\upnu}
=
e^{-\uplambda}
&=&
1-\frac{r_{\scalebox{.56}{\textsc{hor}}}}{r}
\left(1+\frac{Q^2}{r_{\scalebox{.56}{\textsc{hor}}}^2}-\frac{\upalpha\,M}{r_{\scalebox{.56}{\textsc{hor}}}}\,e^{-r_{\scalebox{.56}{\textsc{hor}}}/M}\right)
\nonumber
\\
&&
\qquad\qquad\qquad+\frac{Q^2}{r^2}-\frac{\upalpha\,M}{r}\,e^{-r/M}.
\end{eqnarray}

\blt{To find analytical solutions to $r_{\scalebox{.56}{\textsc{hor}}}$, appropriate values
of the parameters $\upalpha$, $Q$, and $\ell$ must be chosen.
However, since the dominant energy conditions demand $r_{\scalebox{.56}{\textsc{hor}}}\geq 2\,M$,
the choice of these values cannot be arbitrary. 
Evaluating the effective density~\eqref{dendom} at the event horizon, and making use of Eq. \eqref{horizondom} imply that  
\begin{equation}
\label{condi}
Q^2
\geq
\frac{4\upalpha M^2}{e^2}, \qquad
\quad
\ell
\geq
\frac{M}{e^2}.
\end{equation}
The physical interpretation of $Q$ encompasses the case of an electric charge, but not only restricted to it, but also encoding the possibility of hidden gauge charges, tidal charge, and eventually, Kaluza--Klein stringy effects \cite{daRocha:2017cxu}, or any other source. In the case where $Q$ represents an electric charge, the electrovacuum generated by the Reissner--Nordstr\"om solution additionally accommodates a tensor-vacuum that is proportional to $\upalpha$ in Eq.~\eqref{dendom}.
One must emphasize that the Reissner--Nordstr\"{o}m metric has an event horizon 
\begin{equation}
\label{domhor}
r_{{\scalebox{.56}{\textsc{RN}}}}
=
M+\sqrt{M^2-Q^2}
<
2\,M,
\end{equation}
and also an inner Cauchy horizon, 
\begin{equation}
\label{cauchy}
r_{\scalebox{.56}{\textsc{C}}}
\equiv
M-\sqrt{M^2-Q^2}
<
r_{\scalebox{.56}{\textsc{hor}}}.
\end{equation}
The solution~\eqref{dominantBH} can thus yield three ramifications wherein the event horizon $r_{\scalebox{.56}{\textsc{hor}}}$ has straightforward analytical formul\ae\, and the dominant energy conditions are satisfied.
Similarly to the Reissner--Nordstr\"{o}m solution, the three cases to be studied have an inner Cauchy horizon $r_{\scalebox{.56}{\textsc{C}}}<r_{\scalebox{.56}{\textsc{hor}}}$. }

\blt{If the event horizon is made proportional to the mass, as $r_{\scalebox{.56}{\textsc{hor}}}=k\,M$,
as long as $k\ge 2$, to satisfy the dominant energy conditions, the metric components are given by \begin{equation}
\label{dominantBHf}
e^{\nu}
=
e^{-\lambda}
=
1-\frac{2{\cal M}}{r}
+\frac{Q^2}{r^2}
-\frac{\upalpha r_{\scalebox{.56}{\textsc{hor}}}}{kr}e^{-kr/r_{\scalebox{.56}{\textsc{hor}}}},
\end{equation}
where the Reissner--Nordstr\"om-like event horizon reads 
\begin{equation}
\label{domhorf}
r_{\scalebox{.56}{\textsc{hor}}}
=
\tilde{\cal M}
+\sqrt{\tilde{\cal M}^2-\tilde{Q}^2},
\end{equation}
where $\tilde{\cal M}={\cal M}/\upbeta$ and $\tilde{Q}^2={Q}^2/\upbeta$ with
\begin{equation}
\upbeta=1-\upalpha\,\frac{e^{-k}}{k}
\ .
\end{equation}
Therefore the metric components~\eqref{dominantBHf} describe a black hole solution arising from nonlinear electrodynamics, with event horizon 
\begin{eqnarray}
r_{\scalebox{.56}{\textsc{hor}}}
=
\frac{r_{\scalebox{.56}{\textsc{RN}}}}{\upbeta}
\geq
{r_{\scalebox{.56}{\textsc{RN}}}},
\end{eqnarray}
as $\upbeta\leq 1$. The nonlinear electrodynamics is obtained when one identifies
\begin{eqnarray}
\label{tmunu}
\Uptheta_{\mu\nu}
=
-\mathcal{L}(F)g_{\mu\nu}-\mathcal{L}_{F}F_{\mu}^{\;\rho}F_{\rho\nu},
\end{eqnarray}
where 
\begin{equation}
F
=
\frac{1}{4}F_{\rho\sigma}F^{\rho\sigma},
\quad
\quad 
\mathcal{L}_{\scalebox{.56}{\textsc{$F$}}}=\frac{\partial\mathcal{L}}{\partial F}.
\end{equation}
When static, spherically symmetric, stellar distributions described by the metric (\ref{metric}) are regarded, 
the field strength reads 
\begin{equation}
F_{\mu\nu}(r)
=
\left(\updelta^0_\mu\updelta^1_\nu-\updelta^1_\mu\updelta^0_\nu\right)E(r),
\end{equation}
where the electric field is given by 
\begin{eqnarray}
E(r)
=
\frac{Q}{r^2}
-\frac{\upalpha\, e^{-\frac{kr}{r_{\scalebox{.56}{\textsc{hor}}}}} }
{4r_{\scalebox{.56}{\textsc{hor}}}Q}(kr+2r_{\scalebox{.56}{\textsc{hor}}}).
\end{eqnarray}
Introducing the field 
\begin{equation}
P
=\mathcal{L}_{\scalebox{.56}{\textsc{$F$}}}^2 F_{\rho\sigma}F^{\rho\sigma},
\end{equation}
the underlying nonlinear electrodynamics can be thus placed into the
$P$-dual framework \cite{Salazar:1987ap,Ovalle:2017fgl}, described by the Lagrangian 
\begin{eqnarray}
\label{Lp}
\!\!\!\!\!\mathcal{L}(P)
\!=\!
-4 \pi P
\!-\!\frac{\upalpha k({-2P})^{\frac14}}
{4\sqrt{\pi Q} \,r_{\scalebox{.56}{\textsc{hor}}}}\exp\left[\frac{k\sqrt{Q}}{2\, \sqrt{\pi }({-2P})^{\frac14}r_{\scalebox{.56}{\textsc{hor}}}}\right],
\end{eqnarray}
where
\begin{eqnarray}
\mathfrak{G}(P)
=
-\frac{k\sqrt{Q}}{2\, \sqrt{\pi }r_{\scalebox{.56}{\textsc{hor}}}({-2P})^{\frac14}}.
\end{eqnarray}
One can read off the relation $Q^2\sim\upalpha$,
yielding the Schwarzschild standard solution whenever $\upalpha\to 0$.
When $Q$ represents an electric charge, one can state that the Reissner--Nordstr\"om electrovacuum is permeated by a tensor-vacuum governed by \eqref{Lp}. }

\par
As concrete examples, one can saturate the inequalities~\eqref{condi},
and \eqref{dominantBHH} become, defining
$
\mathsf{M}=M\left(1+\frac{\upalpha}{2\,e^2}\right)$, 
\begin{equation}
\label{98}
e^{\upnu_{\scalebox{.62}{(I)}}}
=
e^{-\uplambda_{\scalebox{.62}{(I)}}}
=
1-\frac{2\mathsf{M}}{r}
+\frac{Q^2}{r^2}
-\frac{\sqrt{\upalpha}\,Q}{2\,r}\,e^{1-2\sqrt{\upalpha}\,r/e\,Q}
,
\end{equation}
\blt{which can be interpreted as a nonlinear electrodynamics coupled with gravity, similarly to the content in the last paragraph} \cite{Ovalle:2020kpd}. 
The event horizons are placed at $r_{\scalebox{.56}{\textsc{hor}}}=2M$, and $
r_{\scalebox{.56}{\textsc{hor}}}
=
\frac{e}{\sqrt{\upalpha}}\,Q.$

In the second case, the relation 
\begin{equation}
\label{qcase2}
Q^2
=
\upalpha\,\ell\,M
\left(2+\upalpha\,e^{-{\upalpha\,\ell}/{M}}\right)
,
\end{equation}
leads to
\begin{eqnarray}
\label{100}
e^{\upnu_{\scalebox{.62}{(II)}}}
=
e^{-\uplambda_{\scalebox{.62}{(II)}}}
&=&
1-\frac{2M+\upalpha\,\ell}{r}
+\frac{2\,\upalpha\,\ell\,M}{r^2}
\nonumber
\\
&&
-\frac{\upalpha\,M}{r^2}\,e^{-r/M}\left(r-\upalpha\,\ell\,e^{\frac{r-\upalpha\,\ell}{M}}\right)
.
\end{eqnarray}
The event horizon is now at $r_{\scalebox{.56}{\textsc{hor}}}=\upalpha\,\ell\geq 2M$.
As $\upalpha\,\ell\sim\,M$, Eq. \eqref{100} can be also realized as a solution in nonlinear electrodynamics coupled with gravity.  

Finally, when 
\begin{equation}
Q^2
=
\upalpha\,M\left(2M+\upalpha\,\ell\right)
e^{-\frac{(2M+\upalpha\,\ell)}{M}}
,
\end{equation}
the metric components $e^{\upnu_{\scalebox{.62}{(III)}}}
=
e^{-\uplambda_{\scalebox{.62}{(III)}}}
$ read
\begin{eqnarray}
\label{102}
e^{\upnu_{\scalebox{.62}{(III)}}}
&=&1
-\frac{\upalpha M}{r^2}\,e^{-r/M}\left[r-\left(2M+\upalpha\ell\right)e^{\frac{r-\left(2M+\upalpha\ell\right)}{M}}\right]
\nonumber
\\
&&\qquad\qquad\qquad\qquad\qquad-\frac{2M+\upalpha\,\ell}{r}.
\end{eqnarray}
The event horizon is at $r_{\scalebox{.56}{\textsc{hor}}}=2M+\upalpha\,\ell=2{\cal M}\geq 2M$.

\section{Weyl and trace anomalies of gravitational decoupled hairy black holes}
\label{dsfb}
\par The holographic Weyl anomaly is reminiscent of the regularization process applied to the gravitational part of the action under conformal transformations \cite{Henningson:1998gx}. In general, the anomaly can be expressed by  
\begin{equation}
\mathcal{A}\propto E_{\left(d\right)}+I_{\left(d\right)},
\end{equation}
where $E_{(d)}$ is the $d$-dimensional Euler density, and $I_{(d)}$ denotes a conformal invariant\footnote{For $d=4$ this invariant is unique, being given by the contraction of the Weyl tensor to itself \cite{Henningson:1998gx}.}. In four dimensions the Euler density takes the form
\begin{equation}
	E_{\left(4\right)}=\frac{1}{64}\left(K-4R^{\mu\nu}R_{\mu\nu}+R^{2}\right).
\end{equation}
Up to a multiplicative constant, the holographic Weyl anomaly becomes
\begin{equation}\label{TCFT}
	\left\langle T\right\rangle _{\scalebox{.62}{CFT}}\sim \left(R^{\mu\nu}R_{\mu\nu}-\frac{1}{3}R^{2}\right).
\end{equation}

\par On the other hand, the trace anomaly is a function of the matter content on a curved background together with its geometric aspects \cite{Birrell:1982ix} \begin{equation} \label{T4D}
	\left\langle T\right\rangle _{\scalebox{.62}{4D}}\sim \left({K}-R^{\mu\nu}R_{\mu\nu}-\Box R\right).
\end{equation}
This anomaly quantifies the deviation from conformal invariance, i. e., the vanishing of this particular quantity indicates that the associated dual theory preserves conformal symmetry.

Trace anomalies from the field theory side can be compared to the one found in the CFT using the coefficient \cite{Casadio:2003jc}
\begin{equation}\label{gammaCFT}
	\Upgamma=\left|1-\frac{\left\langle T\right\rangle _{\scalebox{.62}{4D}}}{\left\langle T\right\rangle _{\scalebox{.62}{CFT}}}\right|,
\end{equation}
where the definitions given in Eqs. (\ref{TCFT}, \ref{T4D}) have been applied. 
The quantity $K = R_{\mu\nu\rho\sigma}R^{\mu\nu\rho\sigma}$ denotes the Kretschmann scalar ${K}$. 
This result holds in the context of asymptotic AdS backgrounds \cite{Henningson:1998gx} and here we discuss the possibility of emulating this result to the gravitational decoupling of hairy black holes. 

\par The coefficient \eqref{gammaCFT} can quantify how AdS/CFT is reliable in the context where the metrics (\ref{98}, \ref{100}, \ref{102}) of gravitational decoupled hairy black holes are taken into account. It measures the trace anomalies associated with them and how the additional sources backreact, in the gravitational decoupling setup. The coefficient $\Upgamma$ can formally run from $0$ to infinity, and AdS/CFT can underlie this setup 
for values that are close to unit \cite{Casadio:2003jc,Meert:2020sqv}.
In fact, for the Schwarzschild case, $\left\langle T\right\rangle _{\scalebox{.62}{4D}}\propto K$ and $\left\langle T\right\rangle _{\scalebox{.62}{CFT}}=0$ yielding $\Upgamma\to\infty$, what compromises AdS/CFT in the general-relativistic case \cite{Casadio:2003jc}. 
\par On braneworld scenarios, one seeks for spacetimes where $\Upgamma \ll 1$, where the quantum conformal field theory on the brane and the classical gravity on the bulk descriptions are dual and equivalent \cite{Casadio:2003jc,Shiromizu:2001ve}. The case when $\Upgamma=1$ was obtained evaluating the coefficient \eqref{gammaCFT} for a braneworld black hole solution \cite{Casadio:2003jc}. One expects other solutions to be more intricate, as in the case of gravitational decoupled hairy black holes (\ref{98}, \ref{100}, \ref{102}). 

\begin{figure}[h]
\centering
	\includegraphics[width=8.6cm]{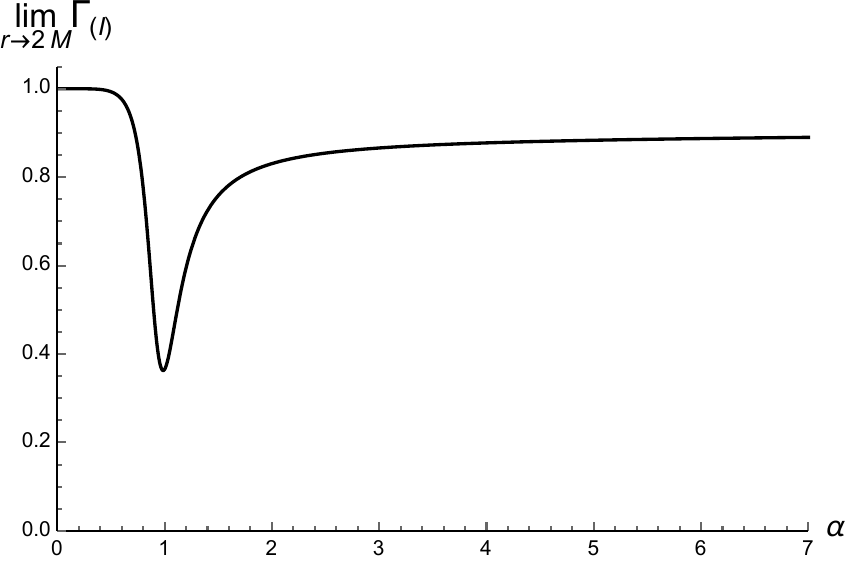}
\caption{Plot of $\lim_{r\to2M}\Upgamma_{\scalebox{.62}{(I)}}$ as a function of the decoupling parameter $\upalpha$.}
\label{f1}
\end{figure}

\begin{figure}[h]
\centering
	\includegraphics[width=8.6cm]{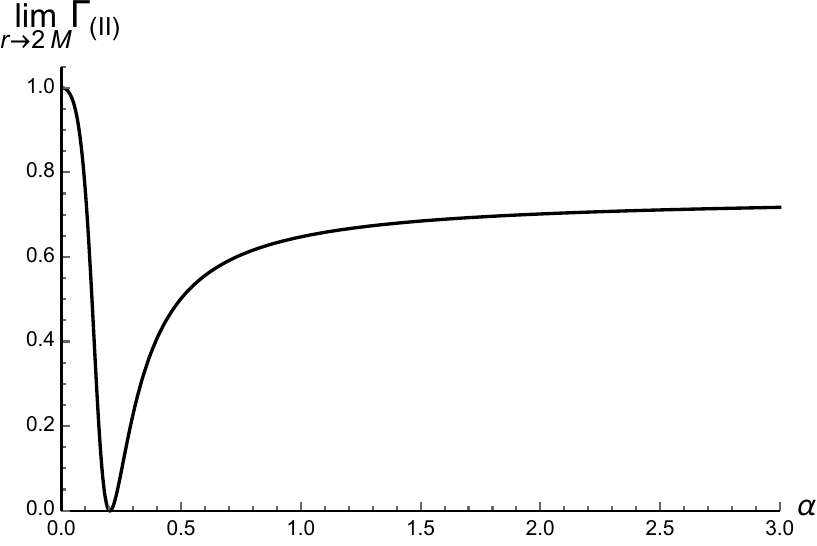}
\caption{Plot of $\lim_{r\to2M}\Upgamma_{\scalebox{.62}{(II)}}$ as a function of the decoupling parameter $\upalpha$.}
\label{f2}
\end{figure}

\begin{figure}[H]
\centering
	\includegraphics[width=8.6cm]{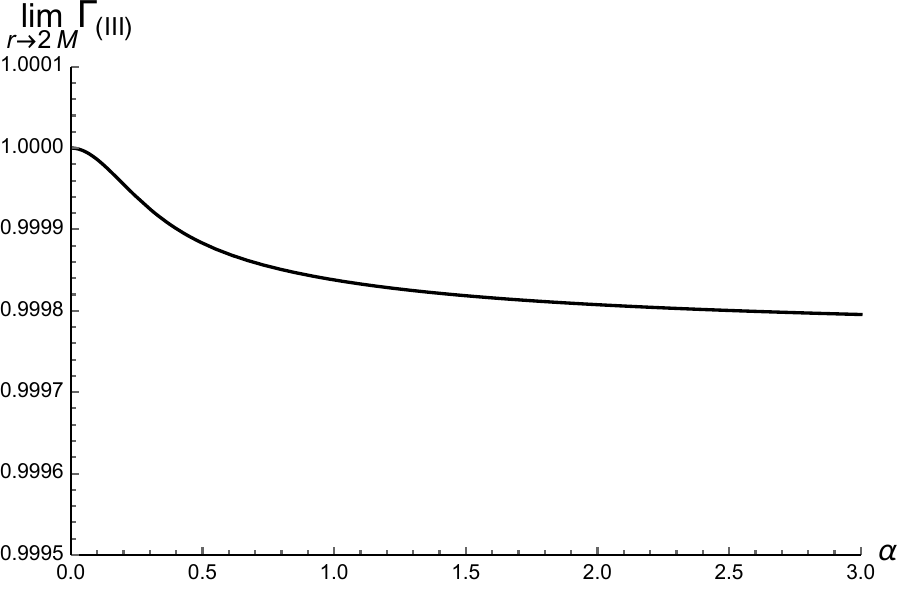}
\caption{Plot of $\lim_{r\to2M}\Upgamma_{\scalebox{.62}{(III)}}$ as a function of the decoupling parameter $\upalpha$.}
\label{f3}
\end{figure}

Notice that for large values of $r$, all three coefficients associated with  (\ref{98}, \ref{100}, \ref{102}) have near unit values, 
\beq
\lim_{r\to\infty}\Upgamma^{\scalebox{.63}{(I)}}\approx 1,\\
\lim_{r\to\infty}\Upgamma^{\scalebox{.63}{(II)}}\approx 1,\\
\lim_{r\to\infty}\Upgamma^{\scalebox{.63}{(III)}}\approx 1.\eeq
This is interesting and relevant, as such behavior is different of the Schwarzschild kernel metric used to derive these solutions using the gravitational decoupling method. Next, the limiting expressions for $r\to2M$ are analyzed in Figs. \ref{f1} -- \ref{f3}, which display $\Upgamma|_{r\to 2M}$ as a function of $\upalpha$. Since $\upalpha$ is not a perturbation parameter, at least technically it can assume any value. Thus the coefficient $\Upgamma_{\scalebox{.62}{CFT}}$ can be determined in the $\upalpha \to \infty$ limit, 
\beq\label{g1}
\lim_{\substack{\upalpha\to\infty \\ r \to 2M}}\Upgamma_{\scalebox{.62}{(I)}} &=& 0.923-\mathcal{O}\left(\frac{1}{\upalpha}\right), \\\label{g2}
 \lim_{\substack{\upalpha\to\infty \\ r \to 2M}}\Upgamma_{\scalebox{.62}{(II)}}&=&1+\mathcal{O}\left(\frac{1}{\upalpha}\right), \\ \label{g3}\lim_{\substack{\upalpha\to\infty \\ r \to 2M}}\Upgamma_{\scalebox{.62}{(III)}}&=&0.\eeq
Considering these limits along with the plots in Figs. \ref{f1} -- \ref{f3}, one can safely state that $\Upgamma\leq 1$ at the regions of interest, namely, for the concomitant limits $r\to \infty$ and $r \to 2M$.

\par In Figs. \ref{f1} and \ref{f2} one can clearly see a bump, where the value of $\Upgamma$ is minimum. These values are $\Upgamma_{\scalebox{.62}{(I)}} \approx 0.367$, when $\upalpha \approx 0.988$, as $\Upgamma_{\scalebox{.62}{(II)}} \approx 0$, for $\upalpha \approx 0.202$. In fact, the tiny range $0.2015 < \upalpha < 0.2032$ yields $\Upgamma_{\scalebox{.62}{(II)}} < 3\times 10^{-5}$.


\par As the limiting values of $\Upgamma$ for $r\to \infty$ do not depend on the parameter $\upalpha$, we can use the values mentioned above on the metrics to conclude that, if the AdS/CFT correspondence holds for these particular solutions obtained via gravitational decoupling, then the best agreement between classical gravity and the associated field theory is then implemented.

\section{Relation to AdS/CFT correspondence} \label{newSect}
\par \blt{Having these results, it is important to point out how the solutions can be used in the context of AdS/CFT correspondence, given the compliance between gravity and the associated boundary theory from the computation of $\Upgamma$. For this connection to be established, one recalls Einstein's equations in its full form
\begin{equation} \label{n1}
	R_{\mu\nu}-\frac{1}{2}Rg_{\mu\nu}=\Lambda_{4}g_{\mu\nu}+\overline{T}_{\mu\nu}, \\
\end{equation}
where $\Lambda_{4}$ is the cosmological constant in 4-dimensions, and $\overline{T}_{\mu\nu}$ is the energy momentum tensor describing all other fields except the cosmological constant one. In Sec. \ref{Sgd} the solutions studied were derived requiring that $T_{\mu\nu}=0$ and $\Theta_{\mu\nu}\neq0$. From the explicit expressions of the solutions, c.f. Eqs. (\ref{98}, \ref{100}, \ref{102}), one can check that the limit $r\mapsto\infty$ leads to the Minkowski spacetime. However, the MGD has a strong connection to brane-world scenarios \cite{Ovalle:2009xk}, which can be employed to establish the connection.}

\par \blt{In the brane-world setup, the stress-energy-momentum tensor associated to the brane has the most general form \cite{GCGR}
\begin{equation}\label{n2}
	\overline{T}_{\mu\nu}=T_{\mu\nu}+E_{\mu\nu}+\gamma^{-1}S_{\mu\nu}+L_{\mu\nu}+P_{\mu\nu},
\end{equation}
where $\gamma$ denotes the brane tension. 
The first term is the ordinary energy-momentum tensor from Einstein's equations, as already pointed out $T_{\mu\nu}=0$. The third term contains corrections of second order on the energy-momentum tensor $T_{\mu\nu}$, such that $S_{\mu\nu}\propto\mathcal{O}\left(T_{\mu\nu}^{2}\right)$, and therefore vanishes as well. The remaining terms in Eq. \eqref{n2} carry non-local effects and also affect the energy-momentum tensor depending on the geometric procedure one uses to embed the brane in the bulk. Specifically, $L_{\mu\nu}$ accounts for the embedding, and is associated with the bending of the brane concerning the codimension-1 bulk. $P_{\mu\nu}$ contains possible stringy effects living in the bulk. $E_{\mu\nu}$ describes a Weyl fluid in the bulk and is responsible for non-local effects and anisotropies. Explicit expressions for these terms can be found in \cite{GCGR,maartens,Antoniadis2011}.
}

\par \blt{Notice that this description is valid on a context where the solutions describe a brane, which is embedded in a higher dimensional space-time, and therefore the quantities appearing on \eqref{n2} are related to the higher dimensional bulk. It is important to remark that the bulk is governed by its own Einstein's equations, such as Eq. \eqref{n1} in one extra dimension, where other matter fields can be defined, and it has its own cosmological constant. The cosmological constants in the bulk and on the brane are related to each other by fine tuning with the brane tension $\gamma$, \cite{Randall:1999ee}
\begin{equation}\label{n:3}
\Lambda_{4}=\frac{\kappa_{5}^{2}}{2}\left(\Lambda_{5}+\frac{1}{6}\kappa_{5}^{2}\gamma^{2}\right),
\end{equation}
where $\kappa_{5}=8\pi G_5$ and $G_5$ is the Newton constant in five dimensions, which is related to the 4 dimensional Newton constant by the Planck length $\ell_p$ as $G_5=\ell_p G_4$. The brane tension cannot be arbitrarily defined and has a predicted lower bound for its value $\gamma\geq2.8131\times10^{-6}$ \cite{Fernandes-Silva:2019fez}. To prevent matter fields living in the bulk to interact with matter fields in the brane one has to fine tune the cosmological constant in Eq. \eqref{n:3} such that $\Lambda_{4}=0$ \cite{Ovalle:2013vna}. Given the lower bound on the brane tension, one finds immediately that
\begin{equation}
	\Lambda_{5}=-\frac{1}{6}\kappa_{5}^{2}\gamma^{2}\\ ,
\end{equation}
Implying that the bulk where the brane is located is an AdS space-time. Considering the overall setup described in Sec. \ref{Sgd}, one can therefore identify $\overline{T}_{\mu\nu}=\alpha\Theta_{\mu\nu}$, therefore
\begin{equation} \label{n:4}
	\alpha\Theta_{\mu\nu}=E_{\mu\nu}+L_{\mu\nu}+P_{\mu\nu}.
\end{equation}
From this connection, the AdS/CFT conjecture can be applied to the described metrics following the prescription of \cite{Soda:2010si}. Ref. \cite{Ovalle:2017fgl} established the way how the general gravitational decoupling can be led to the membrane paradigm of AdS/CFT.}

\section{Conclusions}
\label{4}
The gravitational decoupling of hairy black holes was utilized and inspected with the apparatus provided by trace and Weyl anomalies. The gravitational decoupling was shown to be a trustworthy model, in the context of AdS/CFT. Since the value of the $\Upgamma$ coefficient, for the gravitational decoupling case, was shown to be near the unit, it means that the gravitational decoupling solutions may occupy a privileged place and can play a prominent role in emulating AdS/CFT on gravitational decoupled solutions. 
The $\upalpha\to\infty$ limit in Eqs. (\ref{g1} -- \ref{g3}) can be seen as a regime where the stress-energy-momentum tensor (\ref{emt}) has only the additional source contribution, in the gravitational sector, being the general-relativistic source negligible. This characterizes, in fact, the tensor vacuum regime. The coefficient (\ref{gammaCFT}) quantifies the excitation of gravitationally decoupled matter fields and estimates the signatures of gravitational waves beyond the general-relativistic setup, measuring fluctuations of the decoupled source. Hence, the anomaly coefficient \eqref{gammaCFT} brings useful information about placing gravitational decoupled hairy black holes in the AdS/CFT framework, implementing a method to quantify trace anomalies in this context, besides also quantifying backreaction of gravitationally decoupled additional sources, driven by the parameter $\upalpha$.

\subsection*{Acknowledgements}
PM thanks Coordenação de Aperfeiçoamento de Pessoal de Nível Superior – Brasil (CAPES) – Finance Code 001.
RdR~is grateful to FAPESP (Grants No. 2017/18897-8 and No. 2021/01089-1) and the National Council for Scientific and Technological Development -- CNPq (Grants No. 303390/2019-0, No. 406134/2018-9, and No. 402535/2021-9), for partial financial support.

\bibliography{bib_MGDanomaly}

\providecommand{\newblock}{}
\begin{thebibliography}{10}
\expandafter\ifx\csname url\endcsname\relax
  \def\url#1{{\tt #1}}\fi
\expandafter\ifx\csname urlprefix\endcsname\relax\def\urlprefix{URL }\fi
\providecommand{\eprint}[2][]{\url{#2}}

\bibitem{Ovalle:2017fgl}
Ovalle J 2017 {\em Phys. Rev.\/} {\bf D95} 104019 (\textit{Preprint}
  \eprint{1704.05899})

\bibitem{Ovalle:2019qyi}
Ovalle J 2019 {\em Phys. Lett.\/} {\bf B788} 213--218 (\textit{Preprint}
  \eprint{1812.03000})

\bibitem{Ovalle:2020kpd}
Ovalle J, Casadio R, Contreras E and Sotomayor A 2021 {\em Phys. Dark Univ.\/}
  {\bf 31} 100744 (\textit{Preprint} \eprint{2006.06735})

\bibitem{Casadio:2012rf}
Casadio R and Ovalle J 2014 {\em Gen. Rel. Grav.\/} {\bf 46} 1669
  (\textit{Preprint} \eprint{1212.0409})

\bibitem{Ovalle:2017wqi}
Ovalle J, Casadio R, da~Rocha R and Sotomayor A 2018 {\em Eur. Phys. J.\/} {\bf
  C78} 122 (\textit{Preprint} \eprint{1708.00407})

\bibitem{Antoniadis:1998ig}
Antoniadis I, Arkani-Hamed N, Dimopoulos S and Dvali G~R 1998 {\em Phys.
  Lett.\/} {\bf B436} 257 (\textit{Preprint} \eprint{9804398})

\bibitem{covalle2}
Casadio R and Ovalle J 2014 {\em Gen. Rel. Grav.\/} {\bf 46} 1669
  (\textit{Preprint} \eprint{1212.0409})

\bibitem{Ovalle:2014uwa}
Ovalle J, Gergely L~A and Casadio R 2015 {\em Class. Quant. Grav.\/} {\bf 32}
  045015 (\textit{Preprint} \eprint{1405.0252})

\bibitem{Ovalle:2016pwp}
Ovalle J, Casadio R and Sotomayor A 2017 {\em Adv. High Energy Phys.\/} {\bf
  2017} 9756914 (\textit{Preprint} \eprint{1612.07926})

\bibitem{Casadio:2013uma}
Casadio R, Ovalle J and da~Rocha R 2014 {\em Class. Quant. Grav.\/} {\bf 31}
  045016 (\textit{Preprint} \eprint{1310.5853})

\bibitem{Ovalle:2013vna}
Ovalle J, Linares F, Pasqua A and Sotomayor A 2013 {\em Class. Quant. Grav.\/}
  {\bf 30} 175019 (\textit{Preprint} \eprint{1304.5995})

\bibitem{Ovalle:2013xla}
Ovalle J and Linares F 2013 {\em Phys. Rev.\/} {\bf D88} 104026
  (\textit{Preprint} \eprint{1311.1844})

\bibitem{Ovalle:2018vmg}
Ovalle J and Sotomayor A 2018 {\em Eur. Phys. J. Plus\/} {\bf 133} 428
  (\textit{Preprint} \eprint{1811.01300})

\bibitem{Casadio:2012pu}
Casadio R and Ovalle J 2012 {\em Phys. Lett.\/} {\bf B715} 251--255
  (\textit{Preprint} \eprint{1201.6145})

\bibitem{Casadio:2015jva}
Casadio R, Ovalle J and da~Rocha R 2015 {\em EPL\/} {\bf 110} 40003
  (\textit{Preprint} \eprint{1503.02316})

\bibitem{Casadio:2016aum}
Casadio R and da~Rocha R 2016 {\em Phys. Lett.\/} {\bf B763} 434--438
  (\textit{Preprint} \eprint{1610.01572})

\bibitem{daRocha:2020rda}
da~Rocha R 2020 {\em Symmetry\/} {\bf 12} 508 (\textit{Preprint}
  \eprint{2002.10972})

\bibitem{Fernandes-Silva:2017nec}
Fernandes-Silva A and da~Rocha R 2018 {\em Eur. Phys. J.\/} {\bf C78} 271
  (\textit{Preprint} \eprint{1708.08686})

\bibitem{Contreras:2018gzd}
Contreras E 2018 {\em Eur. Phys. J.\/} {\bf C78} 678 (\textit{Preprint}
  \eprint{1807.03252})

\bibitem{Ovalle:2007bn}
Ovalle J 2008 {\em Mod. Phys. Lett.\/} {\bf A23} 3247--3263 (\textit{Preprint}
  \eprint{gr-qc/0703095})

\bibitem{Sharif:2018tiz}
Sharif M and Saba S 2018 {\em Eur. Phys. J.\/} {\bf C78} 921 (\textit{Preprint}
  \eprint{1811.08112})

\bibitem{Sharif:2019mzv}
Sharif M and Sadiq S 2019 {\em Chin. J. Phys.\/} {\bf 60} 279--289

\bibitem{Morales:2018urp}
Morales E and Tello-Ortiz F 2018 {\em Eur. Phys. J.\/} {\bf C78} 841
  (\textit{Preprint} \eprint{1808.01699})

\bibitem{Rincon:2019jal}
Rinc\'on A, Gabbanelli L, Contreras E and Tello-Ortiz F 2019 {\em Eur. Phys. J.
  C\/} {\bf 79} 873 (\textit{Preprint} \eprint{1909.00500})

\bibitem{Hensh:2019rtb}
Hensh S and Stuchlik Z 2019 {\em Eur. Phys. J. C\/} {\bf 79} 834
  (\textit{Preprint} \eprint{1906.08368})

\bibitem{Ovalle:2019lbs}
Ovalle J, Posada C and Stuchli­k Z 2019 {\em Class. Quant. Grav.\/} {\bf 36}
  205010 (\textit{Preprint} \eprint{1905.12452})

\bibitem{Gabbanelli:2019txr}
Gabbanelli L, Ovalle J, Sotomayor A, Stuchlik Z and Casadio R 2019 {\em Eur.
  Phys. J. C\/} {\bf 79} 486 (\textit{Preprint} \eprint{1905.10162})

\bibitem{Tello-Ortiz2020}
Tello-Ortiz F 2020 {\em Eur. Phys. J.\/} {\bf C80} 413

\bibitem{Gabbanelli:2018bhs}
Gabbanelli L, Rinc\'on A and Rubio C 2018 {\em Eur. Phys. J.\/} {\bf C78} 370
  (\textit{Preprint} \eprint{1802.08000})

\bibitem{Panotopoulos:2018law}
Panotopoulos G and Rinc\'on A 2018 {\em Eur. Phys. J.\/} {\bf C78} 851
  (\textit{Preprint} \eprint{1810.08830})

\bibitem{Heras:2018cpz}
Heras C~L and Leon P 2018 {\em Fortsch. Phys.\/} {\bf 66} 1800036
  (\textit{Preprint} \eprint{1804.06874})

\bibitem{Contreras:2018vph}
Contreras E and Bargue\~no P 2018 {\em Eur. Phys. J.\/} {\bf C78} 558
  (\textit{Preprint} \eprint{1805.10565})

\bibitem{Tello-Ortiz:2020euy}
Tello-Ortiz F, Maurya S and Gomez-Leyton Y 2020 {\em Eur. Phys. J. C\/} {\bf
  80} 324

\bibitem{Fernandes-Silva:2019fez}
Fernandes-Silva A, Ferreira-Martins A~J and da~Rocha R 2019 {\em Phys. Lett.\/}
  {\bf B791} 323--330 (\textit{Preprint} \eprint{1901.07492})

\bibitem{daRocha:2017cxu}
da~Rocha R 2017 {\em Phys. Rev.\/} {\bf D95} 124017 (\textit{Preprint}
  \eprint{1701.00761})

\bibitem{Fernandes-Silva:2018abr}
Fernandes-Silva A, Ferreira-Martins A~J and da~Rocha R 2018 {\em Eur. Phys.
  J.\/} {\bf C78} 631 (\textit{Preprint} \eprint{1803.03336})

\bibitem{Maurya:2019kzu}
Maurya S, Maharaj S and Deb D 2019 {\em Eur. Phys. J. C\/} {\bf 79} 170

\bibitem{PerezGraterol:2018eut}
P\'erez~Graterol R 2018 {\em Eur. Phys. J. Plus\/} {\bf 133} 244

\bibitem{Morales:2018nmq}
Morales E and Tello-Ortiz F 2018 {\em Eur. Phys. J.\/} {\bf C78} 618
  (\textit{Preprint} \eprint{1805.00592})

\bibitem{Contreras:2019iwm}
Contreras E, Rinc\'on A and Bargue\~no P 2019 {\em Eur. Phys. J.\/} {\bf C79}
  216 (\textit{Preprint} \eprint{1902.02033})

\bibitem{Contreras:2019fbk}
Contreras E 2019 {\em Class. Quant. Grav.\/} {\bf 36} 095004 (\textit{Preprint}
  \eprint{1901.00231})

\bibitem{Singh:2019ktp}
Singh K, Maurya S, Jasim M and Rahaman F 2019 {\em Eur. Phys. J. C\/} {\bf 79}
  851

\bibitem{Tello-Ortiz:2019gcl}
Tello-Ortiz F, Maurya S~K, Errehymy A, Singh K and Daoud M 2019 {\em Eur. Phys.
  J.\/} {\bf C79} 885

\bibitem{Maurya:2019xcx}
Maurya S and Tello-Ortiz F 2020 {\em Phys. Dark Univ.\/} {\bf 29} 100577
  (\textit{Preprint} \eprint{1907.13456})

\bibitem{Cedeno:2019qkf}
Linares~Cedeno F~X and Contreras E 2020 {\em Phys. Dark Univ.\/} {\bf 28}
  100543 (\textit{Preprint} \eprint{1907.04892})

\bibitem{Sharif:2018toc}
Sharif M and Sadiq S 2018 {\em Eur. Phys. J.\/} {\bf C78} 410
  (\textit{Preprint} \eprint{1804.09616})

\bibitem{Estrada:2018zbh}
Estrada M and Tello-Ortiz F 2018 {\em Eur. Phys. J. Plus\/} {\bf 133} 453
  (\textit{Preprint} \eprint{1803.02344})

\bibitem{Torres:2019mee}
Torres-Sanchez V and Contreras E 2019 {\em Eur. Phys. J. C\/} {\bf 79} 829
  (\textit{Preprint} \eprint{1908.08194})

\bibitem{Abellan:2020jjl}
Abell\'an G, Rincon A, Fuenmayor E and Contreras E 2020  (\textit{Preprint}
  \eprint{2001.07961})

\bibitem{Estrada:2018vrl}
Estrada M and Prado R 2019 {\em Eur. Phys. J. Plus\/} {\bf 134} 168
  (\textit{Preprint} \eprint{1809.03591})

\bibitem{Leon:2019abq}
León P and Sotomayor A 2019 {\em Fortsch. Phys.\/} {\bf 67} 1900077
  (\textit{Preprint} \eprint{1907.11763})

\bibitem{Casadio:2019usg}
Casadio R, Contreras E, Ovalle J, Sotomayor A and Stuchlick Z 2019 {\em Eur.
  Phys. J. C\/} {\bf 79} 826 (\textit{Preprint} \eprint{1909.01902})

\bibitem{Sharif:2019mjn}
Sharif M and Waseem A 2019 {\em Annals Phys.\/} {\bf 405} 14--28

\bibitem{Abellan:2020wjw}
Abellan G, Torres-Sanchez V, Fuenmayor E and Contreras E 2020 {\em Eur. Phys.
  J. C\/} {\bf 80} 177 (\textit{Preprint} \eprint{2001.08573})

\bibitem{Rincon:2020izv}
Rinc\'on A, Contreras E, Tello-Ortiz F, Bargueno P and Abell\'an G 2020 {\em
  Eur. Phys. J. C\/} {\bf 80} 490 (\textit{Preprint} \eprint{2005.10991})

\bibitem{Sharif:2020arn}
Sharif M and Majid A 2020 {\em Phys. Dark Univ.\/} {\bf 30} 100610
  (\textit{Preprint} \eprint{2006.04578})

\bibitem{Maurya:2020gjw}
Maurya S, Singh K~N and Dayanandan B 2020 {\em Eur. Phys. J. C\/} {\bf 80} 448

\bibitem{Capper:1973mv}
Capper D and Duff M 1974 {\em Nucl. Phys. B\/} {\bf 82} 147

\bibitem{Duff:1993wm}
Duff M 1994 {\em Class. Quant. Grav.\/} {\bf 11} 1387--1404 (\textit{Preprint}
  \eprint{hep-th/9308075})

\bibitem{Henningson:1998gx}
Henningson M and Skenderis K 1998 {\em JHEP\/} {\bf 07} 023 (\textit{Preprint}
  \eprint{hep-th/9806087})

\bibitem{Kuntz:2017pjd}
Kuntz I 2018 {\em Eur. Phys. J. C\/} {\bf 78} 3 (\textit{Preprint}
  \eprint{1712.06582})

\bibitem{Bonora:1983ff}
Bonora L, Cotta-Ramusino P and Reina C 1983 {\em Phys. Lett. B\/} {\bf 126}
  305--308

\bibitem{Bonora:2014qla}
Bonora L, Giaccari S and Lima~de Souza B 2014 {\em JHEP\/} {\bf 07} 117
  (\textit{Preprint} \eprint{1403.2606})

\bibitem{Kuntz:2019omq}
Kuntz I and da~Rocha R 2020 {\em Nucl. Phys. B\/} {\bf 961} 115265
  (\textit{Preprint} \eprint{1909.10121})

\bibitem{Casadio:2003jc}
Casadio R 2004 {\em Phys. Rev. D\/} {\bf 69} 084025 (\textit{Preprint}
  \eprint{hep-th/0302171})

\bibitem{Ovalle:2018umz}
Ovalle J, Casadio R, da~Rocha R, Sotomayor A and Stuchlik Z 2018 {\em Eur.
  Phys. J.\/} {\bf C78} 960 (\textit{Preprint} \eprint{1804.03468})

\bibitem{Salazar:1987ap}
Salazar I~H, Garcia A and Plebanski J 1987 {\em J. Math. Phys.\/} {\bf 28}
  2171--2181

\bibitem{Birrell:1982ix}
Birrell N and Davies P 1984 {\em {Quantum Fields in Curved Space}\/} Cambridge
  Monographs on Mathematical Physics (Cambridge, UK: Cambridge Univ. Press)
  ISBN 978-0-521-27858-4, 978-0-521-27858-4

\bibitem{Meert:2020sqv}
Meert P and da~Rocha R 2021 {\em Nucl. Phys. B\/} {\bf 967} 115420
  (\textit{Preprint} \eprint{2006.02564})

\bibitem{Shiromizu:2001ve}
Shiromizu T, Torii T and Ida D 2002 {\em JHEP\/} {\bf 03} 007
  (\textit{Preprint} \eprint{hep-th/0105256})

\bibitem{Ovalle:2009xk}
Ovalle J 2009 {Braneworld Stars: Anisotropy Minimally Projected Onto the Brane}
  {\em {9th Asia-Pacific International Conference on Gravitation and
  Astrophysics (ICGA 9) Wuhan, China, June 28-July 2, 2009}\/} pp 173--182
  (\textit{Preprint} \eprint{0909.0531})

\bibitem{GCGR}
Shiromizu T, Maeda K~i and Sasaki M 2000 {\em Phys. Rev.\/} {\bf D62} 024012

\bibitem{maartens}
Maartens R and Koyama K 2010 {\em Living Rev. Rel.\/} {\bf 13} 5

\bibitem{Antoniadis2011}
Antoniadis I, Atkins M and Calmet X 2011 {\em JHEP\/} {\bf 2011} 1387

\bibitem{Randall:1999ee}
Randall L and Sundrum R 1999 {\em Phys. Rev. Lett.\/} {\bf 83} 3370--3373
  (\textit{Preprint} \eprint{hep-ph/9905221})

\bibitem{Soda:2010si}
Soda J 2011 {\em Lect. Notes Phys.\/} {\bf 828} 235--270 (\textit{Preprint}
  \eprint{1001.1011})

\end{thebibliography}

\end{document}